\documentclass[10pt]{iopart}

\usepackage{graphicx}
\usepackage{bm}
\usepackage{amsmath}
\usepackage[caption=false]{subfig}
\newcommand{\phantomsubfloat}[1]{
    {
        \captionsetup[subfigure]{labelformat=empty}
        \subfloat[][]{#1}
    }%
}
\begin{document}

\title[Core transport changes in DIII-D discharges with off-axis Te profile peaks]{Investigation of core transport changes in DIII-D discharges with off-axis Te profile peaks}

\author{R Xie$^1$, M E Austin$^2$, K Gentle$^2$ and C C Petty$^3$}
\address{ 
$^1$University of Wisconsin-Madison, Madison, WI 53706, USA
}%

\address{ 
$^2$Institute for Fusion Studies, The University of Texas at Austin, Austin, TX 78712, USA
}%

\address{ 
$^3$General Atomics, San Diego, CA 92121, USA
}%
\ead{rxie9@wisc.edu}

\begin{abstract}
DIII-D discharges that transition to H-mode solely with off-axis electron cyclotron heating (ECH) often exhibit strong off-axis peaking of electron temperature profiles at the heating location. Electron heat transport properties near these off-axis temperature peaks have been studied using modulated ECH. The Fourier analyzed electron temperature data have been used to infer electron thermal diffusivity. Comparisons with numerical solutions of the time-dependent electron thermal equation find that the data are consistent with a narrow region with electron diffusivity $\chi_e$ an order of magnitude lower than the average value across the plasma, suggesting an electron internal transport barrier (ITB) near the ECH heating location. Detailed profile analysis and equilibrium reconstructions suggest that the formation of these ITBs are correlated with off-axis values of the safety factor $q$ being near 1. Furthermore, the ECH driven H-mode discharges demonstrate more rapid electron heating rate near the ECH deposition location than L-mode discharges with higher auxiliary ECH heating power. Additional modeling attributes this difference to the modification of electron heat transport in the core at the L-H transition, which also sustains the off-axis electron temperature peaks.
\end{abstract}

%
\vspace{2pc}
%
\submitto{\PPCF}
%
%
\ioptwocol

\section{\label{sec:intro}Introduction}

Plasma transport dictates the degree of plasma confinement in tokamaks, and thus is a critical factor in achieving fusion power \cite{hazeltine_book}. In order to improve plasma confinement and increase fusion gain, high confinement modes (H-modes) \cite{Wagner2007}, which are characterized by a pedestal region with reduced plasma transport near the plasma edge, have been explored and utilized. 

Due to the strong dependence of energy confinement and fusion performance on pedestal parameters \cite{Doyle2007}, many studies have been dedicated to the investigation of pedestal transport. On the other hand, the need for better understanding of transport mechanisms in the core remains. One of these mechanisms is the change in heat, momentum or particle transport induced by auxiliary heating such as electron cyclotron heating (ECH). Previous experiments performed on the Rijnhuizen Tokamak Project (RTP) tokamak with dominant off-axis ECH observed significantly hollow $T_e$ profile \cite{Hogeweij_1996, Cardozo_1997}. These unusual $T_e$ profiles periodically form sharp ears, i.e., prominent off-axis maxima with large $T_e$ gradients on both sides of the peak. In addition, a recent study in the Large Helical Device (LHD) stellarator has observed quasi-steady-state hollow $T_e$ profiles when heating with off-axis ECH \cite{tsujimura2022}. These experimental results highlight a change in core plasma transport during the application of off-axis ECH.



Regarding the phenomenon observed on the RTP tokamak, previous modelings have linked electron thermal diffusivity $\chi_e$ to safety factor $q$ and suggested that the hollow $T_e$ profiles form when ECH is deposited precisely on top of an internal transport barrier (ITB) \cite{Wolf2003} located near a low order rational $q$ surface \cite{Cardozo_1997, Hogeweij_1998,de_baar_99, Schilham_2001}. Further simulation effort confirmed the presence of negative convective heat flux in the core, which sustains the observed hollow profiles \cite{DeBaar2005, Mantica2005}. An outward heat convection that sustains the hollow $T_e$ profile is also observed in the LHD experiment \cite{tsujimura2022}, but is not linked to the rotational transform profiles. While some questions regarding hollow $T_e$ profiles have been answered for the L-mode case, similar phenomena and modification of core plasma transport in ECH dominant H-mode discharges have not been closely examined.


A dedicated experiment was performed on DIII-D to study steady state hollow $T_e$ profiles with sharp gradient changes in H-mode discharges. Transport analysis highlights the presence of a region with reduced plasma transport, which is the characteristic of an ITB, near the plasma core during off-axis $T_e$ peaking. Simulation efforts using a linear transport model also indicate a difference in core heat transport between L-mode and H-mode discharges. The present work aims to provide quantitative experimental data in order to promote theoretical investigation into the formation and stability of these unusual profiles. The experimental setup and results for this dedicated experiment are presented in section \ref{sec:EXP}. Comparisons with the transport model is shown in section \ref{sec:GEVOL}. Finally, a conclusion will be given in section \ref{sec:Sum}.

\section{\label{sec:EXP}Dedicated experiment}

\begin{figure}[!t]
\phantomsubfloat{\label{fig:bata}}
\phantomsubfloat{\label{fig:batb}}
\phantomsubfloat{\label{fig:batc}}
\includegraphics{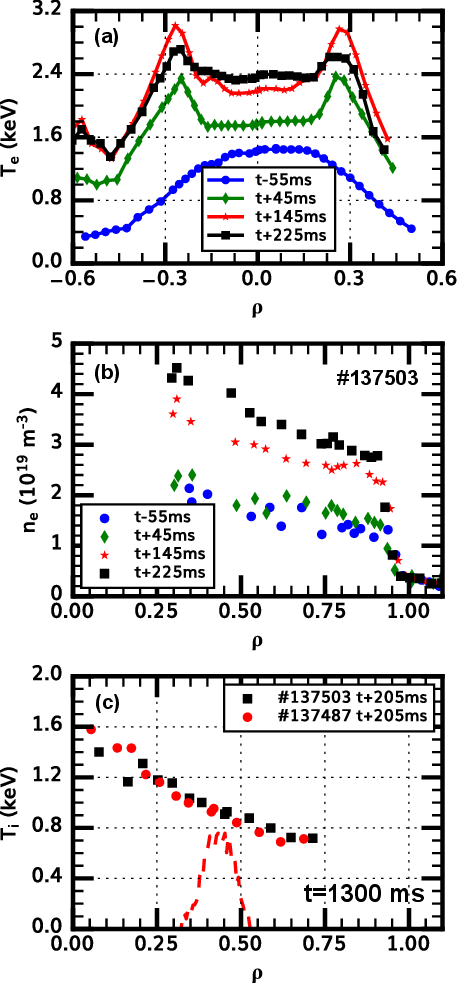}
\caption{\label{fig:H_batear} Samples of (a) $T_e$ profile measured by ECE, (b) $n_e$ profile measured by Thomson scattering, and (c) $T_i$ profiles measured by CER in off-axis ECH heated H-mode DIII-D discharges. The negative $\rho$ values in (a) indicate the high field side of magnetic axis. $T_i$ profile for discharge 137487 in (c) is measured when strong off-axis $T_e$ peak is present. The steady-state ECH deposition profile (red dashed line, in a.u.) is also shown in (c). The experimental data clearly shows sustained off-axis peaking of $T_e$ after ECH is applied at $t=1300$ ms close to $\rho \sim 0.4$ on high field side.}
\end{figure}

A dedicated experiment was performed on the DIII-D tokamak to study the unusual off-axis electron temperature peaks observed in ECH dominant H-mode discharges (see figure \ref{fig:bata}). Previous studies \cite{Petty_1994, Smith_2015} on electron energy transport during off-axis ECH experiments were performed in L-mode. This experiment focuses on the ELM-free phase after a purely ECH-driven L to H-mode transition (figure \ref{fig:H_history}). The plasma is in a lower single null configuration with elongation $\kappa=1.8$ and average triangularity $\delta\sim0.5$. The line averaged electron density is low during the initial ohmic phase, $\bar{n}_{e}\approx 1.9\times 10^{19}\  \mathrm{m^{-3}}$. The toroidal magnetic field is set to $B_t = 1.8$ T with plasma current $I_P = 0.75$ MA, the major radius is $R_0=1.68$ m, and the minor radius is $a=0.61$ m. The resulting safety factor is $q_{95} \sim 6.7$, and the normalized plasma pressure is $\beta_N \sim 1.1$ after H-mode transition. Five gyrotrons were used to launch 2.8 MW of ECH power at 110 GHz with X-mode polarization. In some discharges, one gyrotron was modulated, resulting in the ECH power cycling between 2.1 MW and 2.7 MW. While the neutral beam injection (NBI) heating power was not zero, only short 10 ms NBI ``blips'' every 100 ms were used for diagnostic purposes. These short pulses did not provide substantial heating or alter the plasma states significantly. 

The radial $n_e$ and $T_e$ profiles are measured with Thomson scattering  \cite{TS_ref} every $12.5$ ms. $T_e$ is also measured with an electron cyclotron emission (ECE) radiometer \cite{ECE_ref} at a higher time resolution every $0.2$ ms. The carbon impurity ion temperature, density and toroidal rotation velocity are measured with charge exchange recombination (CER) spectroscopy \cite{CER_ref}. The ion parameters are also calculated using linear inter- and extrapolation of the available CER measurements. The TORAY-GA ray-tracing code \cite{TORAY_ref} was used to determine the ECH power deposition profiles and deposition location $\rho_{dep}$ (normalized flux coordinate). Equilibrium reconstructions are created with the EFIT code \cite{Lao_1985_EFIT} utilizing external magnetics data and kinetic profile constraints from TRANSP modeling \cite{TRANSP_ref,TRANSP_OMFIT_ref}. Internal motional Stark effect (MSE) \cite{MSE_ref} data is not always included as a constraint since the measurement is only available during NBI diagnostic ``blips''. The safety factor $q$ profiles are then obtained from these reconstructions.

\subsection{\label{ssec:Hmode}Observation of off-axis electron temperature peaks in ECH heated H-mode discharges}

\begin{figure}
\phantomsubfloat{\label{fig:2a}}
\phantomsubfloat{\label{fig:2b}}
\phantomsubfloat{\label{fig:2c}}
\phantomsubfloat{\label{fig:2d}}
\includegraphics{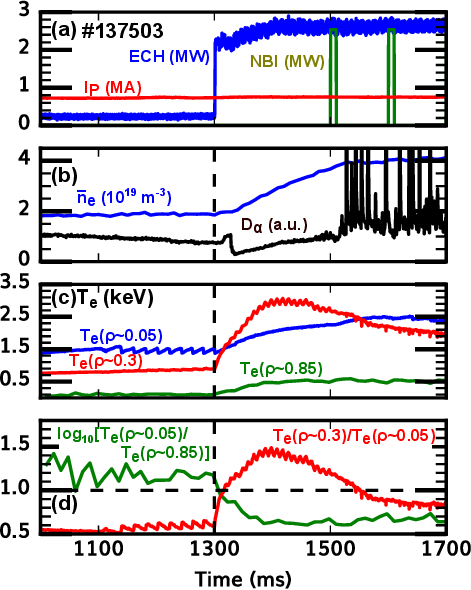}
\caption{\label{fig:H_history} Time history plot of (a) injected neutral beam, ECH power and plasma current $I_P$, (b) line averaged electron density $\bar{n}_e$ and a $D_\alpha$ signal, (c) $T_e$ measured by ECE near the peak (red, $\rho\sim 0.3$), at the core (blue, $\rho\sim 0.05$), and by Thomson scattering on top of the pedestal (green, $\rho\sim 0.85$), and (d) ratios of $T_e$ measurements in (c) for one of the H-mode discharges with modulated ECH.}
\end{figure}

When the ECH is applied at 1300 ms near $\rho_{dep}\sim0.4$, an L-H transition occurs within 50 ms, indicated by the drop in $D_\alpha$ emission intensity in figure \ref{fig:2b}. Figure \ref{fig:2c} shows the evolution of electron temperature measured by ECE near the magnetic axis ($\rho \sim 0.05$), near the peak ($\rho\sim 0.3$), and by Thomson scattering on top of the pedestal ($\rho\sim 0.85$). The first $10$ ms diagnostic NBI pulse is injected 200 ms (300 ms in some discharges) after ECH application to avoid the effect of NBI heating on plasma states. The $T_i$ profile (figure \ref{fig:batc}) is centrally peaked at this time. The $T_e$ ratio between off-axis peak and core ($\frac{T_e(\rho\sim 0.3)}{T_e(\rho\sim 0.05)}$) as well as $T_e$ ratio between core and pedestal ($\log_{10}\frac{T_e(\rho\sim 0.05)}{T_e(\rho\sim 0.85)}$) are shown in figure \ref{fig:2d}. When ECH is turned on, electron temperature grows faster near $\rho_{dep}$ than inside $\rho\sim0.23$, which is the approximate sawtooth inversion radius observed by ECE during the ohmic phase. Consequently, the ratio between off-axis peak and core in figure \ref{fig:2d} quickly increases and becomes larger than unity, indicating the presence of an off-axis electron temperature peak. The line-averaged electron density (figure \ref{fig:2b}) increases on a time scale similar to the core electron temperature. Both electron temperature and density reach equilibrium approximately 250 ms after ECH turns on. The $T_e$ ratio between off-axis peak and core (figure \ref{fig:2d}) also decreases below unity around this time. The $T_e$ ratio between core and pedestal initial drops due to the formation of $T_e$ pedestal, then remains relatively constant afterwards. In figure \ref{fig:2c}, we observe similar $T_e$ growth rates at the core ($\rho\sim0.05$) and on top of the pedestal ($\rho\sim0.85$) while $T_e$ growth rate near the off-axis peak ($\rho\sim0.3$) is much higher. This difference suggests that the growth of the off-axis $T_e$ peak is not caused by the formation of the pedestal alone.

\begin{figure}
\includegraphics{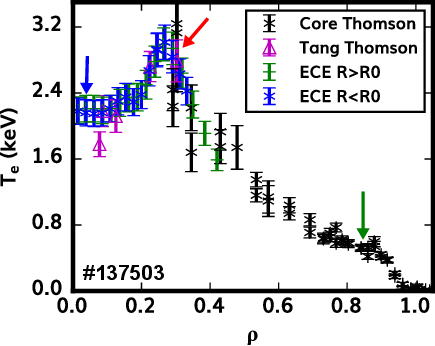}
\caption{\label{fig:non_thermal} Radial profile of $T_e$ at $1445$ ms, $145$ ms after ECH is applied. $T_e$ measurements by Thomson scattering and ECE show the off-axis peak at the same location with matching magnitude. The locations of ECE and Thomson scattering channels shown in figure \ref{fig:2c} are marked with arrows of the same color.}
\end{figure}

Since high power radio-frequency (RF) heating can produce non-maxwellian electron distribution functions ($f_e$) that distort the ECE signal near the ECH deposition location \cite{Harvey_non_thermal,Subhas_non_thermal}, it is fair to ask if the off-axis peaks observed in figure \ref{fig:bata} could be caused by non-thermal high energy electrons created by the applied ECH. To confirm the thermal nature of the ECE electron temperature measurements, we compare the experimental data from ECE and Thomson scattering. Figure \ref{fig:non_thermal} shows the electron temperature measured by ECE and Thomson scattering at a time slice when the off-axis $T_e$ peak is observed. In this example, Thomson scattering observes an off-axis peak around the same $\rho$ value with similar magnitude as the ECE data. $T_e$ from Thomson scattering and ECE also match on both sides of the peak. In addition, the electron temperature peak is clearly observed on both sides of the magnetic axis with roughly symmetrical location and magnitude in figure \ref{fig:bata}. If $f_e$ were non-Maxwellian, it would generally be expected to exhibit a strong asymmetry in ECE-Te profile measurements \cite{Harvey_non_thermal}. Since no asymmetry is observed, and there is good agreement with Thomson scattering, the electron distribution function is presumed to be Maxwellian.

\begin{figure}
\includegraphics{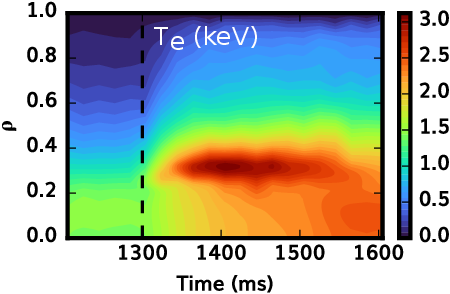}
\caption{\label{fig:H_2d_batear} 2D plot of the fitted (versus $\rho$) electron temperature from $1205$ ms to $1605$ ms. The vertical dashed line indicate the time ECH is turned on ($1300$ ms). ECH is injected close to $\rho\sim 0.4$.}
\end{figure}

Figure \ref{fig:bata} and \ref{fig:H_2d_batear} show the evolution of electron temperature profiles before and after the ECH is turned on at $t=1300$ ms. The positive and negative $\rho$ values in figure \ref{fig:bata} indicate the low field side (LFS) and the high field side (HFS) of the magnetic axis, respectively. At $t-35$ ms, before the ECH is applied, we can see an electron temperature profile peaked inside $\rho=0.1$. Approximately $20$ ms after the ECH is applied, the previously mentioned off-axis electron temperature peak can be observed. Figure \ref{fig:H_2d_batear} shows that the largest change in electron temperature after ECH turns on happens around $\rho=0.3$ while $T_e$ change is smaller inside $\rho=0.2$; $T_e$ then continues to grow over the entire radial profile until approximately $t+100$ ms, when the peak $T_e$ starts to saturate and remains relatively stationary for the next $\sim60$ ms; $T_e$ inside $\rho = 0.2$ continues to grow. It should be noted that the location of off-axis $T_e$ peak ($\rho\sim 0.3$) is different from the ECH deposition location $\rho_{dep}\sim0.4$. This behavior is different from the RTP \cite{Hogeweij_1996} and LHD \cite{tsujimura2022} experiments, where the off-axis $T_e$ peaks coincide with $\rho_{dep}$. 

The plasma enters a grassy ELM phase at around $t+160$ ms, and $T_e$ near $\rho=0.3$ starts to gradually decay. At $t+225$ ms, before the first major ELM event at $t+227$ ms, the off-axis peak is still observed in figure \ref{fig:bata}. It is worth-noting that the 3 outermost ECE channels on the LFS are experiencing density cutoff, so the ECE measurements outside $\rho=0.3$ at $t+225$ ms in figure \ref{fig:bata} are likely lower than the actual $T_e$ values. After the ELM event at $t+227$ ms, the $T_e$ profile becomes flattened inside $\rho=0.3$ and the off-axis peaks are no longer clearly visible. Nevertheless, the off-axis $T_e$ peak lasts $\approx 200$ ms, which is much longer than the electron energy confinement time $\tau_{Ee}\approx 60$ ms.

\begin{figure}
\phantomsubfloat{\label{fig:da}}
\phantomsubfloat{\label{fig:db}}
\phantomsubfloat{\label{fig:dc}}
\phantomsubfloat{\label{fig:dd}}
\includegraphics{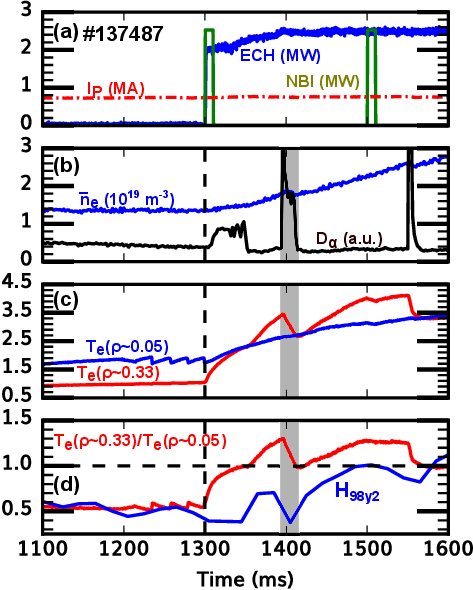}
\caption{\label{fig:dither_history} Time history plot of (a) injected neutral beam, ECH power and plasma current $I_P$, b) line averaged electron density $\bar{n}_e$ and a $D_\alpha$ signal, (c) $T_e$ measured by ECE near the peak (red, $\rho\sim 0.33$) to $T_e$ at the core (blue, $\rho\sim 0.05$), (d) ratio of the two $T_e$ measurements in (c) and the calculated $H_{98y2}$ confinement factor for the dithering H-mode discharge. This discharge briefly dropped out of H-mode inside the shaded region.}
\end{figure}

Figure \ref{fig:dither_history} shows the time history of experimental parameters for a discharge with intermittent H-mode transition. In this rare case, the plasma briefly transitioned back into L-mode between $1395$ ms and $1415$ ms. The $D_\alpha$ emission intensity in figure \ref{fig:db} shows a major event at approximately $1400$ ms; the calculated $H_{98y2}$ factor, which is the normalized energy confinement time with respect to the $\tau_{E,98y2}$ scaling \cite{Doyle2007}, also decreases to $0.4$ in figure \ref{fig:dd}, which is lower than during the ohmic phase. As the discharge back transitions to L-mode, the growing off-axis peak quickly disappears; in figure \ref{fig:dc} and figure \ref{fig:dd}, we can see that $T_e$ near $\rho\sim 0.3$ quickly decreases below $T_e$ inside the core and their ratio becomes less than unity between the two vertical dotted lines. Immediately after H-mode is recovered around $1415$ ms, $T_e$ near $\rho \sim 0.3$ increases and the off-axis peak is observed until the ELM event at $1550$ ms. It should also be noted that this discharge first entered H-mode at a later time ($\sim1350$ ms) than the discharge shown in Fig. \ref{fig:H_history} ($\sim 1330$ ms). Due to these delays, the diagnostic NBI injection at $1500$ ms is able to measure $T_i$ when the off-axis $T_e$ peak is fully developed (figure \ref{fig:batc}, red circles). There is no obvious difference compared to the $T_i$ profile (figure \ref{fig:batc}, black squares) when the off-axis $T_e$ peak is decaying.

\subsection{\label{ssec:Lmode}L-mode comparison}

\begin{figure*}
\phantomsubfloat{\label{fig:la}}
\phantomsubfloat{\label{fig:lb}}
\phantomsubfloat{\label{fig:lc}}
\phantomsubfloat{\label{fig:ld}}
\phantomsubfloat{\label{fig:le}}
\phantomsubfloat{\label{fig:lf}}
\includegraphics{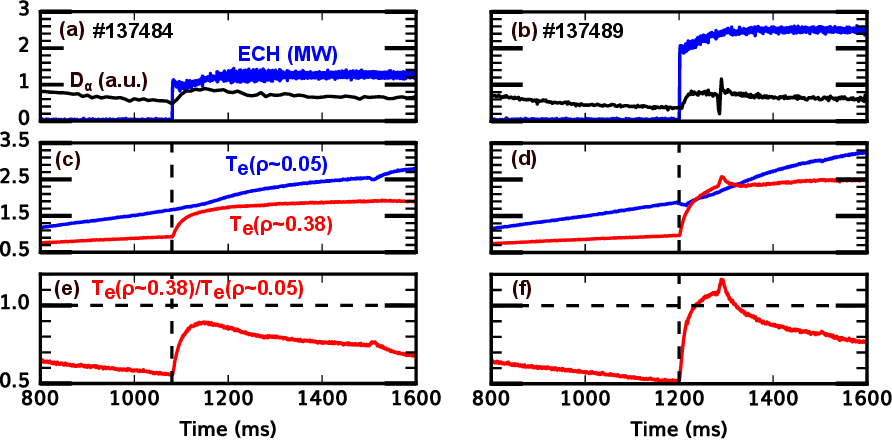}
\caption{\label{fig:L_history} (a) and (b) Time history plot of ECH power and a $D_\alpha$ signal for a pair of L-mode discharges with different ECH heating power. (c) and (d) $T_e$ measured by ECE channels close to the ECH deposition location (red, $\rho\sim0.38$) and inside the core (blue, $\rho\sim0.05$). (e) and (f) Ratio between the $T_e$ measurements in (c) and (d). }
\end{figure*}

The experiment is also performed in L-mode for comparison. In these discharges, the lower triangularity is increased from $0.5$ to $0.6$, and the average electron density $\bar{n}_{e}$ during the ohmic phase is decreased from $1.9\times 10^{19}\  \mathrm{m^{-3}}$ to $1.4\times 10^{19}\  \mathrm{m^{-3}}$. These changes raise the L-H threshold, causing the discharge to remain in L-mode with up to 2.8 MW of ECH power. Figure \ref{fig:L_history} shows a pair of L-mode discharges with $1.5$ and $2.8$ MW of ECH heating power. In both cases, electron density remains the same after ECH switch-on. No off-axis peak is observed during the low power L-mode discharge; the ratio in figure \ref{fig:le} stays less than unity. It is worth-noting that an hollow $T_e$ profile is observed in the high power L-mode discharge. However, the electron heating rate near $\rho_{dep}$ is slower than that in H-mode discharges with lower auxiliary ECH power. No central $T_e$ cooling was observed during the formation of hollow $T_e$ profiles (Fig. \ref{fig:ld}), which is a marked difference from previous RTP experiments \cite{Hogeweij_1996}. Power balance analysis from ONETWO \cite{ONETWO_ref} also shows that heat transfer from electrons to ions does not increase after ECH switch-on. As a result, the hollow $T_e$ profile is not due to additional energy sink near the magnetic axis.

In figure \ref{fig:ld} we can see that $T_e$ near $\rho_{dep}$ grows larger than $T_e$ in the core. The ratio between the two is shown in figure \ref{fig:lf} and becomes greater than unity at around $1235$ ms until $1320$ ms. Near the off-axis $T_e$ peak, the normalized electron collisionality $\nu_e^*\ll 1$ and is comparable to the $\nu_e^*$ in the H-mode discharges. However, this off-axis peak is much less prominent and persists for less time than its counterpart in H-mode discharges; as in figure \ref{fig:2d}, the ratio increases to approximately $1.5$ and is greater than unity for more than $200$ ms. The difference in off-axis $T_e$ peak evolution between L-mode and H-mode discharges with similar $\nu_e^*$ suggests that collisionality is unlikely the driving force of core transport changes. 


It should be noted that we observe a brief L-H transition in L-mode discharges with more than $2$ MW of ECH during this campaign. For example, the discharge shown in Fig. \ref{fig:lb} briefly entered H-mode around $1285$ ms, indicated by the oscillation in $D_\alpha$ emission intensity (see figure \ref{fig:lb}). This transition caused a quick rise in $T_e$ near $\rho_{dep}\sim 0.40$ (figure \ref{fig:ld}) at the same time and extended the duration of off-axis peak shown in Fig. \ref{fig:lf}. 

A similar behavior can be observed during the intermittent H-mode discharge shown in figure \ref{fig:dither_history}. In figure \ref{fig:dc}, we can clearly see an increase in the $T_e$ growth rate near $\rho\sim 0.33$ after the first L-H transition around $1350$ ms while the rate of increase of $T_e$ near the magnetic axis ($\rho\sim 0.05$) remains the same. This change in $T_e$ response only has a delay time of a few milliseconds or less compared to the drop in $D_\alpha$ emission intensity, which indicates that there is a fast reduction in local transport. From these observations, we hypothesize that the growth of the off-axis $T_e$ peak is the result of local transport changes in the core after the L-H transition at the edge. Qualitatively, this suggests that the prominence and extended lifetime of the off-axis peak is related to L-H transition.

\subsection{\label{ssec:ITB-observation}Observation of internal transport barrier}

The sustained steep electron temperature gradient $\nabla T_e$ on both sides of the off-axis peak is a signature of reduced electron heat transport and suggests that an ITB has formed in the region. To confirm the presence of the ITB, we compare changes of $T_e$ profile induced by perturbations before and after the off-axis peak disappears. Since the time scale of these changes is faster than the acquisition rate of Thomson scattering measurements, this section focuses on the experimental $T_e$ data from the ECE radiometer.

\begin{figure*}
\phantomsubfloat{\label{fig:elma}}
\phantomsubfloat{\label{fig:elmb}}
\includegraphics{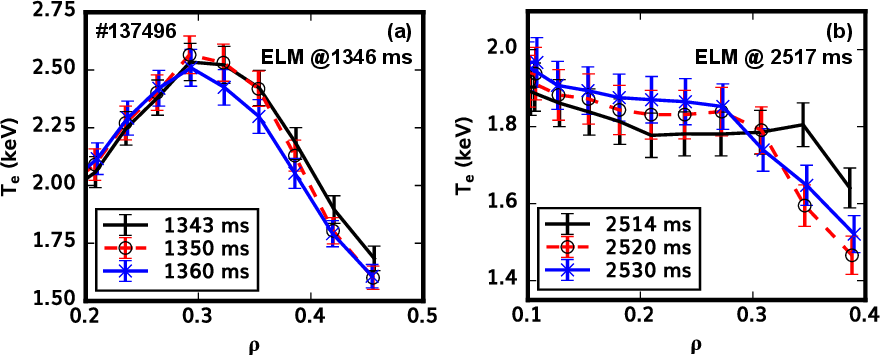}
\caption{\label{fig:496_ELM} Radial $T_e$ profiles measured by ECE before (no symbol, black solid line), less than $5$ ms after (circle, red dashed line) and more than $10$ ms after ($\times$, blue solid line) an ELM event (a) with and (b) without clear off-axis $T_e$ peak. The ELM events are at $1346$ ms and $2517$ ms.}
\end{figure*}

Two types of perturbative heat pulses are analyzed: cold pulses induced by ELM events and modulated ECH (MECH). Figure \ref{fig:496_ELM} shows the impact of ELM event on $T_e$ profiles (a) with and (b) without the off-axis peak. In figure \ref{fig:elma}, notice that the $T_e$ profile inside $\rho\sim 0.35$ at $1350$ ms remains the same as $1343$ ms, before the ELM event. From $1350$ to $1360$ ms, only $T_e$ between $\rho\sim 0.3$ and $0.4$ has visibly decreased. It is evident that there is a delay in $T_e$ drop approaching the $T_e$ peak at $\rho\sim 0.3$. Compare figure \ref{fig:elmb} with figure \ref{fig:elma}: ECE channels outside $\rho\sim 0.3$ in figure \ref{fig:elmb} show similar $T_e$ reduction during the ELM event without any visible delay. This observation suggests that the cold pulse caused by the ELM event is damped as it propagates through the region with high $\nabla T_e$. Qualitatively, this supports the existence of an ITB in the region surrounding the off-axis $T_e$ peak.


It is possible to quantitatively infer the presence of this ITB from ECE $T_e$ data in discharges with MECH. One discharge with MECH is shown in figure \ref{fig:H_history}. In this example, the MECH is deposited off-axis near $\rho\sim 0.3$ and varies slightly during the shot due to electron density evolution. The gyrotron is modulated at $100$ Hz as a square wave with $50\%$ duty cycle ($\Delta t=10$ ms). The applied MECH periodically provides a localized heat deposition to electrons, creating an oscillation in $T_e$ without affecting $n_e$. The $T_e$ time traces are analyzed with standard fast Fourier transform (FFT) techniques, and we can observe the propagation of the heat pulses.

\begin{figure}
\phantomsubfloat{\label{fig:mecha}}
\phantomsubfloat{\label{fig:mechb}}
\includegraphics{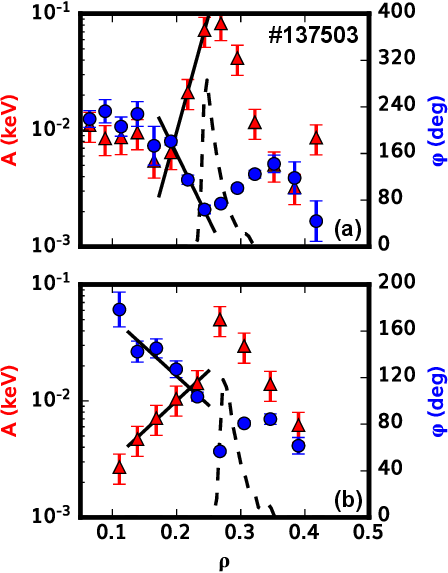}
\caption{\label{fig:H_mech} Radial profile of first harmonic MECH amplitude (red triangles) and phase (blue circles) when (a) the off-axis $T_e$ peak is present ($1313-1383$ ms) and (b) after the off-axis peak has disappeared ($3080-3180$ ms). The solid black lines represent the linearly fits of $\phi$ and $\mathrm{ln}(A)$. The MECH deposition profiles calculated with TORAY are plotted in dashed lines. The change in MECH deposition location is due to electron density evolution.}
\end{figure}


Figure \ref{fig:H_mech} shows the extracted amplitude ($A$) and phase lag ($\varphi$) profiles of the heat pulse at the modulation frequency ($f=100$ Hz). The signal to noise ratio for FFT profiles at higher harmonics is much lower than the first harmonic, and the amplitude profiles do not show clear peaks. Nevertheless, we can compare the first harmonic FFT profiles during different time intervals of the same discharge. The first FFT time interval ($1313-1383$ ms), during which the off-axis $T_e$ peak is present, is limited to $7$ modulation cycles to avoid the effect of drifting ECE channel (figure \ref{fig:mecha}). A longer FFT time interval ($3080-3180$ ms) of $10$ cycles is used when the plasma is in steady state and $T_e$ is peaked inside the core (figure \ref{fig:mechb}). 


In figures \ref{fig:mecha} and \ref{fig:mechb}, the FFT amplitude is the highest and the phase is the lowest at the same position, which agrees with the calculated MECH deposition location. This confirms the estimated MECH deposition profiles calculated with TORAY-GA. Moreover, it allows us to assess that there is no significant heat convection in this region. In the presence of strong heat convection, which was observed in previous RTP results \cite{Mantica2000,Mantica2005}, the first harmonic amplitude peak will shift inward or outward relative to the MECH deposition location. In contrast, the amplitude and phase profiles shown in figure \ref{fig:H_mech} have characteristics of diffusive transport: the amplitude decreases and phase increases away from the deposition location. 




In cylindrical geometry, the heat pulse diffusivity $\chi_e^{HP}$ can be calculated using the amplitude $A$ and phase $\varphi$ with \cite{jacchia_cylinder_chi}:
\begin{equation}
\label{eq:cylinder_chi}
    \chi_{e}^{HP}=\frac{(3/4)\omega}{-\varphi'(A'/A+1/2r)}
\end{equation}
where $\omega=2\pi f$, and $r$ is the minor radius of the measurement location. $\varphi'$ and $A'/A$ are estimated at $\rho\sim 0.2$ by linearly interpolating $\varphi$ and $\mathrm{ln}(A)$ around the location of interest. The fitted diffusivities at $\rho\sim 0.2$ are $0.16\pm 0.03\ \mathrm{m^{2}/s}$ when the off-axis $T_e$ peak is present (figure \ref{fig:mecha}) and $2.5\pm 0.8\ \mathrm{m^{2}/s}$ at the later time without the off-axis $T_e$ peak (figure \ref{fig:mechb}).

Although the exact values of the diffusion coefficient are uncertain, comparison between the two time intervals accurately determines the relative magnitude of diffusion coefficients. Figure \ref{fig:mecha} shows a sharp drop in $A$ and rise in $\varphi$ on both sides of the off-axis $T_e$ peak, which matches the regions with high $\nabla T_e$ in figure \ref{fig:bata}. The heat pulse becomes strongly damped when it propagates through the ITB. There is also a discontinuity in the $A$ and $\varphi$ slopes near the foot of the off-axis $T_e$ peak at $\rho \sim 0.2$. In contrast, figure \ref{fig:mechb} shows a different behavior: after an initial $A$ drop and $\varphi$ rise, the slopes of $A$ and $\varphi$ profiles are less steep than those in figure \ref{fig:mecha} and have no clear discontinuity. This indicates that, when the off-axis $T_e$ peak is present, there is a region with reduced heat diffusivity between $\rho=0.2$ and $0.4$, which is consistent with the existence of an ITB. 








\section{\label{sec:GEVOL}Initial comparison with transport model}

Motivated by observations in the previous section, we analyze the experimental result using a linear transport modeling code. This simulation effort aims to derive radial diffusivity $\chi_e$ and convection velocity $v_e$ profiles that would reproduce the off-axis $T_e$ peaks and quantify transport changes driven by ECH application. 

The code is constructed to solve the time-dependent electron thermal diffusion equation
\begin{equation}
\label{eq:gevol}
    \frac{3}{2}n_e\frac{dT_e}{dt}+\nabla q_e=S_e
\end{equation}
in cylindrical coordinates. $q_e$ and $S_e$ denote the electron thermal flux and net input power to electron. Radial $S_e$ profiles are obtained using power balance calculations by ONETWO and TORAY-GA, which account for Ohmic and ECH input power, radiated loss and electron-ion energy exchange. In this simulation, $n_e$ is assumed to be constant and only $T_e$ is simulated over time. The electron thermal flux $q_e$ can be described using a simple model
\begin{equation}
\label{eq:qe}
    -q_e=n_e\chi_e\nabla T_e + v_en_eT_e
\end{equation}
where $\chi_e$ and $v_e$ denote the electron heat diffusivity and electron heat convection velocity.

\begin{figure}
\includegraphics{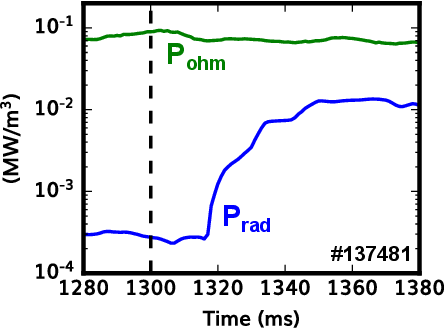}
\caption{\label{fig:Prad_ohm} Time evolution of Ohmic heating power density $P_{ohm}$ and radiated loss $P_{rad}$ calculated with TRANSP near the magnetic axis at $\rho=0.21$ in an H-mode discharge. The vertical dashed line indicate the time ECH is turned on ($1300$ ms).}
\end{figure}

Based on experimental observations presented in section \ref{ssec:Hmode}, we divide the simulation into three phases: ohmic phase, ECH pre H-mode phase and ECH H-mode phase. In the ohmic phase, we assume $v_e=0$ and reproduce the steady-state ohmic $T_e$ profile using a purely diffusive model. After the ECH is turned on, we invoke a heat pinch inside the core. Based on previous RTP modeling works \cite{Schilham_2001,Mantica2005} and the observation that radiated loss is much smaller than the Ohmic heating power density (figure \ref{fig:Prad_ohm}), we expect the existence of an outward heat convection inside the off-axis $T_e$ peak. In addition, the inward shift of $T_e$ maximum relative to $\rho_{dep}$ suggests that heat convection is inward around $\rho_{dep}$. Thus, to model the heat pinch and simulate the effect of convection, we use two parameters $v_{1}$ and $v_{2}$ to create $v_e$ input profiles of the form:
\begin{eqnarray}
\label{eq:ve}
    v_e=
    \begin{cases}
        0, & \rho=0, \\
        v_1, & \rho=\rho_c, \\
        v_{2}, & \text{inside ECH deposition region}, \\
        0, & \text{outside ECH deposition region}.
    \end{cases}
\end{eqnarray}
$\rho_c$ is chosen to be at the foot the off-axis peaks, and the rest of the $v_e$ profile is generated using linear interpolation. The $\chi_e$ profile is also changed in response to the application of ECH. In the ECH pre H-mode phase, $T_e$ starts from the Ohmic profile, grows over the radial profile, and forms the initial off-axis $T_e$ peak. $35$ ms after the ECH is switched on, the simulation enters the ECH H-mode phase. In the third phase, we continue from the pre H-mode off-axis peaked profile and recreate the experimentally observed $T_e$ profile evolution. Both the $\chi_e$ and $v_e$ profiles are modified at L-H transition.

To mimic the effect of ECH, $S_e$ profiles are assumed to be constant except at the start of ECH application. This simplification reduces the free parameters to three $\chi_e$ profiles and two $v_e$ profiles. Using $S_e$, $\chi_e$ and $v_e$ profiles as inputs, the code solves Eq. \ref{eq:gevol} numerically using the Crank–Nicolson method to derive the $T_e$ profiles. The $\chi_e$ and $v_e$ profile inputs are then optimized to minimize the difference between simulated $T_e$ profile and the experimental one.

\subsection{Modeling results}

\begin{figure}
\includegraphics{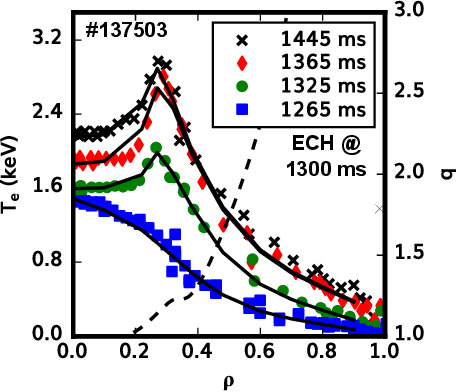}
\caption{\label{fig:gevol_vs} Optimized results of linear simulation. Experimental (symbols) and simulated (solid lines) $T_e$ profiles for H-mode discharge 137503 during ohmic phase (blue squares), during ECH pre H-mode phase (green circles), and after L-H transition (red diamonds and black crosses). ECH is turned on at $1300$ ms. L-H transition occurs at approximately $1335$ ms. Safety factor $q$ profile during off-axis peaking is also shown (dashed line).}
\end{figure}

\begin{figure*}
\phantomsubfloat{\label{fig:gevolDVa}}
\phantomsubfloat{\label{fig:gevolDVb}}
\includegraphics{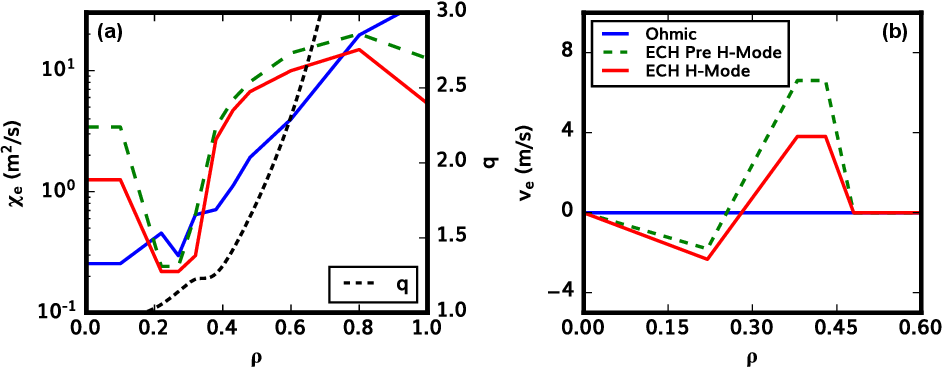}
\caption{\label{fig:gevol_DV} Optimized input profiles of transport coefficients for simulation shown in figure \ref{fig:gevol_vs}. (a) $\chi_e$, and (b) $v_e$ input profiles during ohmic phase (blue solid lines), ECH pre H-Mode phase (green dashed lines), and ECH H-mode phase (red solid lines). Safety factor $q$ profile during off-axis peaking is also shown in (a) (black dotted line).}
\end{figure*}

The simulated and experimental $T_e$ profiles are shown in figure \ref{fig:gevol_vs}. Figure \ref{fig:gevol_DV} shows the associated $\chi_e$ and $v_e$ input profiles that reproduce the off-axis $T_e$ peak evolution in an H-mode discharge. Although this simulation effort recreates the experimental profile, it should be noted that this model has many simplifications and free parameters to uniquely determine the transport coefficients. 

In figure \ref{fig:gevolDVa}, we observe a depression in $\chi_e$ profile around $\rho\sim 0.3$ during ohmic and ECH phases. The location of this $\chi_e$ well during the ohmic phase agrees with the approximate sawtooth inversion radius ($\rho\sim0.23$) observed by ECE, suggesting that it is near the $q=1$ surface. The $\chi_e$ depression deepens after ECH turns on, and becomes widened after L-H transition. 

In figure \ref{fig:gevolDVb}, we observe an inward convection on the outside of the $T_e$ peak and an outward convection on the inside of the $T_e$ peak. The negative $v_e$ represents outward heat convection. This indicates that the convective component in equation \ref{eq:qe} is transporting heat towards the $T_e$ peak against the $T_e$ gradient. The inward convection near $\rho_{dep}$ also shifts the $T_e$ peak towards the magnetic axis, creating a mismatch between the off-axis $T_e$ peak and $\rho_{dep}$. After L-H transition, the inward convection near $\rho_{dep}$ is reduced while the outward convection near $\rho_c$ is slightly increased. This change in net heat convection can be caused by the reduction of inward heat pinch component, the increase of outward convection, or a combination of both. However, we currently do  not have enough evidence to distinguish the two mechanisms from each other.

\begin{figure}
\includegraphics{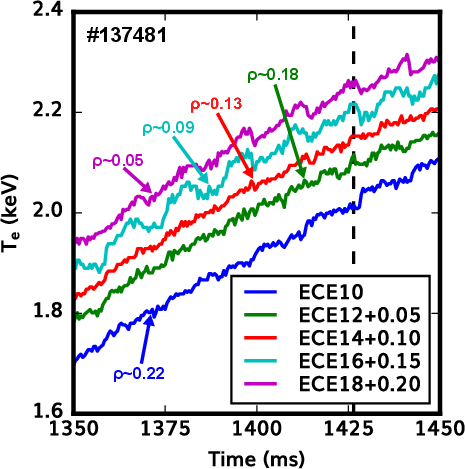}
\caption{\label{fig:sawtooth} Time history plot of $T_e$ that shows sawtooth oscillations during off-axis $T_e$ peaking in an H-mode discharge. The $T_e$ traces, except the bottom one, have been shifted for clarity. The vertical dashed line indicates where the $T_e$ crash and inversion are most noticeable.}
\end{figure}

Figures \ref{fig:gevol_vs} and \ref{fig:gevolDVa} show the $q$ profiles obtained from EFIT equilibrium reconstructed with kinetic pressure and current constraints. In figure \ref{fig:gevolDVa}, we observe the $q=1$ surface near the inside foot of the ITB and a flattened $q$ profile with value close to $6/5$ near the outside edge of the ITB. Figure \ref{fig:sawtooth} shows the time evolution of $T_e$ measured by $5$ ECE channels inside $\rho=0.3$ during off-axis $T_e$ peaking. $T_e$ crashes are observed inside $\rho=0.1$ and inverted sawtooth oscillations are observed at $\rho\sim0.22$. Since $T_e$ profile is flat inside $\rho=0.2$, the sawtooth oscillation amplitude is below the ECE noise level for channels located between $\rho=0.1$ and $\rho=0.2$. Nonetheless, this observation confirms the existence of $q=1$ surface near $\rho\sim 0.2$ during off-axis $T_e$ peaking. 

Previous modeling efforts on off-axis $T_e$ peaks in the RTP tokamak have found similar thermal barrier around a low order rational q surface when using a $q$-comb model, in which the $\chi_e$ profile is a function of $q$ and consists of a series of $\chi_e$ wells centered around low order rational $q$ values \cite{Hogeweij_1998}. It should be noted that the location of off-axis $T_e$ peak coincide with $\rho_{dep}$ and a low order rational $q$ surface in the RTP cases. In contrast, the off-axis peaks are shifted inward relative to $\rho_{dep}$ in this study. This suggests that the ITB observed in H-mode discharge may not be characterized with a single $q$ value; the enhanced ITB is likely a collection of two or more ITBs centered around different low order rational $q$ surfaces between $q=1$ and $q=2$. Unfortunately, internal MSE data is not available as a constraint during these time slices, and the $q$ profiles shown have large uncertainty. It is alternatively possible for $q$ to be around $1$ between $\rho\sim0.2$ and $\rho\sim 0.4$, which means the ITB corresponds to a single low order rational value $q=1$. However, TORAY calculation shows that ECH is deposited at $\rho_{dep}\sim0.4$, and the current drive is localized between $\rho\sim 0.3$ and $\rho\sim 0.5$. This observation led us to incline towards the first hypothesis since we do not expect the $q$ profile to be strongly modified between $\rho\sim0.2$ and $\rho\sim0.3$ without significant current drive. Provisionally, we hypothesize that the presence of low-order rational $q$ surfaces and flattening of $q$ profile inside $\rho_{dep}$ during L-H transition leads to the observed core transport changes in ECH-driven H-mode discharges.

It is noteworthy that $\chi_e$ at $\rho\sim 0.2$ is comparable to the value extracted using FFT analysis in section \ref{ssec:ITB-observation}. The consistency of transport coefficient provides some confidence in the capability of this simulation. Although this model does not include many aspects of ITB physics, it allows us to qualitatively access formation of off-axis $T_e$ peaks and provides further evidence that the experimentally observed off-axis $T_e$ peaks are caused by the presence of ITB and an outward heat convection inside the core.

\begin{figure*}
\phantomsubfloat{\label{fig:relaxa}}
\phantomsubfloat{\label{fig:relaxb}}
\includegraphics{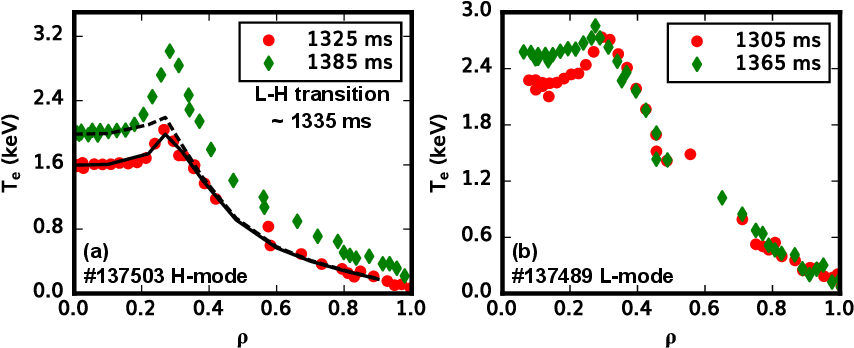}
\caption{\label{fig:L_relax} Comparison of experimental $T_e$ profiles between (a) H-mode and (b) L-mode discharges at two times after the formation of off-axis $T_e$ peaks. In (a), the solid line is the simulation result during the ECH pre H-mode phase, and is the same as the one shown in figure \ref{fig:gevol_vs}. The dashed line is the simulation result at $1385$ ms that uses the ECH pre H-mode $\chi_e$ and $v_e$ input profiles after L-H transition at approximately $1335$ ms.}
\end{figure*}

\subsection{\label{LH_change}Transport changes at L-H transition}
 
Figure \ref{fig:L_relax} compares the evolution of $T_e$ profiles between H-mode and L-mode discharges after the formation of off-axis peak. After the L-H transition, the off-axis $T_e$ peak continues to grow and $T_e$ gradient is strongly negative between $\rho=0.2$ and $\rho=0.3$. In the L-mode discharge, the off-axis $T_e$ peak stop growing and $-\nabla T_e$ changes from negative towards positive.

In the simulation, changes to the $\chi_e$ and $v_e$ input profiles at L-H transition are required to reproduce the experimental $T_e$ behavior. If no modification to the input profiles is made, the simulation fails to capture the growth of off-axis $T_e$ peak. The dashed line in Fig. \ref{fig:relaxa} shows the simulated $T_e$ profile after evolving Eq. \ref{eq:gevol} for an additional $60$ ms using the input profiles from the ECH pre H-mode phase. Even though these input profiles can reproduce the initial formation of off-axis $T_e$ peaks, the $T_e$ profile will relax to a monotonic one. This type of behavior is observed in L-mode discharges with off-axis peaks, as shown in Fig. \ref{fig:relaxb}.

These observations suggests that the sustainment of strongly negative $T_e$ gradient on the inside of the $T_e$ peaks can not be explained with the presence of outward heat convection alone. To reproduce the experimentally observed time evolution of $T_e$ peaks, either a transport change or additional electron heat flux is required inside the core after L-H transition.

\section{\label{sec:Sum}Conclusion}
In summary, unambiguous signatures of core electron heat transport barriers have been seen in DIII-D H-mode discharges triggered solely with electron cyclotron heating.  This is clearly evidenced with observations of enduring hollow $T_e$ profiles and abrupt phase jumps in electron heat pulse analysis in these plasmas. In addition, the difference in electron temperature evolution between L-mode and H-mode cases confirms that the transition to H-mode is a key part of sustaining the off-axis $T_e$ peaks. Comparison with a linear transport model suggests that there is a change in core electron heat transport during L-H transition. This observation is consistent with previous studies \cite{Schmitz_2012,kobayashi_2023} that observe prompt transport change and confinement improvement in the plasma core at L-H transitions. The work has implications for future tokamak devices that intend to reach the H-mode state with ECH as the dominant auxiliary heating method.


The exact mechanisms of ITB formation and transport change in this class of discharge remain unclear. One potential trigger of ITB is the change in $q$ profile. For example, integer $q_{min}$ crossing \cite{Austin2006} and non-monotonic $q$ profile \cite{Osorio2021_ITB_q} can both cause the onset of ITBs, but we are unable to validate this with experimental data due to the lack of MSE diagnostic measurements during L-H transition. Another likely cause of transport change is the flattening of electron density gradient and the reduction of associated density gradient driven turbulence during L-H transition. In the dedicated experiment, the core $T_e$ response to ECH changes extremely fast after L-H transition. These observations are consistent with the suppression of electron micro-instabilities, leading to the improved confinement in the plasma core. This type of stabilization effect has been observed in simulation works of cold-pulse experiments \cite{cold_pulse}.

Furthermore, the strongly negative $T_e$ gradient on both sides of the $T_e$ peaks can be explained with the presence of convective terms in electron thermal transport. This agrees with previous RTP and LHD results \cite{Hogeweij_1996, Mantica2005, tsujimura2022}. However, it is uncertain whether the same mechanisms give rise to the convective counter-gradient transport. The RTP experiments were in L-mode with high density and $\nu_e^*\simeq 1$, whereas $\nu_e^*\ll 1$ in \mbox{DIII-D}. The application of ECH in RTP also had a large impact on plasma current due to its relatively small size. In contrast, the LHD discharges were almost current-less with no significant current change due to ECH.

In the future, it would be good to obtain measurements of turbulent fluctuations in the apparent barrier region in the core that could verify cause and effect of the transport reduction; these were not available in the experiments presented here. Experiments in different parameter spaces and with on-axis ECH would also allow projections of this phenomenon's potential impact on ITER. Additionally, enhancements to turbulent transport modeling codes to be able handle non-standard profiles, like the ones seen in this research, would enable clarification of whether the barrier physics is related specifically to current or density profile physics.

\ack
This material is based upon work supported by the U.S. Department of Energy, Office of Science, Office of Fusion Energy Sciences, using the DIII-D National Fusion Facility, a DOE Office of Science user facility, under Awards DE-FG02-97ER54415 and DE-FC02-04ER54698.

\textbf{Disclaimer:} This report was prepared as an account of work sponsored by an agency of the United States Government. Neither the United States Government nor any agency thereof, nor any of their employees, makes any warranty, express or implied, or assumes any legal liability or responsibility for the accuracy, completeness, or usefulness of any information, apparatus, product, or process disclosed, or represents that its use would not infringe privately owned rights. Reference herein to any specific commercial product, process, or service by trade name, trademark, manufacturer, or otherwise does not necessarily constitute or imply its endorsement, recommendation, or favoring by the United States Government or any agency thereof. The views and opinions of authors expressed herein do not necessarily state or reflect those of the United States Government or any agency thereof.
\section*{References}
\bibliography{ppcf2023}

\providecommand{\noopsort}[1]{}\providecommand{\singleletter}[1]{#1}%
\begin{thebibliography}{10}

\bibitem{hazeltine_book}
R.~D. Hazeltine.
\newblock {\em Plasma confinement}.
\newblock Dover Publications, Mineola, N.Y, 2003 - 1992.

\bibitem{Wagner2007}
F~Wagner.
\newblock A quarter-century of h-mode studies.
\newblock {\em Plasma Physics and Controlled Fusion}, 49(12B):B1, Nov 2007.

\bibitem{Doyle2007}
E.~J. Doyle, W.~A. Houlberg, Y.~Kamada, V.~Mukhovatov, T.~H. Osborne,
  A.~Polevoi, G.~Bateman, J.~W. Connor, J.~G. Cordey, T.~Fujita, X.~Garbet,
  T.~S. Hahm, L.~D. Horton, A.~E. Hubbard, F.~Imbeaux, F.~Jenko, J.~E. Kinsey,
  Y.~Kishimoto, J.~Li, T.~C. Luce, Y.~Martin, M.~Ossipenko, V.~Parail,
  A.~Peeters, T.~L. Rhodes, J.~E. Rice, C.~M. Roach, V.~Rozhansky, F.~Ryter,
  G.~Saibene, R.~Sartori, A.~C.C. Sips, J.~A. Snipes, M.~Sugihara, E.~J.
  Synakowski, H.~Takenaga, T.~Takizuka, K.~Thomsen, M.~R. Wade, and H.~R.
  Wilson.
\newblock Chapter 2: Plasma confinement and transport.
\newblock {\em Nuclear Fusion}, 47, 2007.

\bibitem{Hogeweij_1996}
G.~M.~D. Hogeweij, A.~A.~M. Oomens, C.~J. Barth, M.~N.~A. Beurskens, C.~C. Chu,
  J.~F.~M. van Gelder, J.~Lok, N.~J. Lopes~Cardozo, F.~J. Pijper, R.~W. Polman,
  and J.~H. Rommers.
\newblock Steady-state hollow electron temperature profiles in the rijnhuizen
  tokamak project.
\newblock {\em Phys. Rev. Lett.}, 76:632--635, Jan 1996.

\bibitem{Cardozo_1997}
N~J~Lopes Cardozo, G~M~D Hogeweij, M~de~Baar, C~J Barth, M~N~A Beurskens, F~De
  Luca, A~J~H Donn{\'{e}}, P~Galli, J~F~M van Gelder, G~Gorini, B~de~Groot,
  A~Jacchia, F~A Karelse, J~de~Kloe, O~G Kruijt, J~Lok, P~Mantica, H~J van~der
  Meiden, A~A~M Oomens, Th~Oyevaar, F~J Pijper, R~W Polman, F~Salzedas, F~C
  Schüller, and E~Westerhof.
\newblock Electron thermal transport in {RTP}: filaments, barriers and
  bifurcations.
\newblock {\em Plasma Physics and Controlled Fusion}, 39(12B):B303--B316, dec
  1997.

\bibitem{tsujimura2022}
T.~I. Tsujimura, T.~Kobayashi, K.~Tanaka, K.~Ida, K.~Nagaoka, M.~Yoshinuma,
  I.~Yamada, H.~Funaba, R.~Seki, S.~Satake, T.~Kinoshita, T.~Tokuzawa,
  N.~Kenmochi, H.~Igami, K.~Mukai, M.~Goto, and Y.~Kawamoto.
\newblock Direct observation of the non-locality of non-diffusive
  counter-gradient electron thermal transport during the formation of hollow
  electron-temperature profiles in the large helical device.
\newblock {\em Physics of Plasmas}, 29(3):032504, 2022.

\bibitem{Wolf2003}
R.~C. Wolf.
\newblock Internal transport barriers in tokamak plasmas.
\newblock {\em Plasma Physics and Controlled Fusion}, 45, 2003.

\bibitem{Hogeweij_1998}
G.M.D Hogeweij, N.J.~Lopes Cardozo, M.R.~De Baar, and A.M.R Schilham.
\newblock A model for electron transport barriers in tokamaks, tested against
  experimental data from {RTP}.
\newblock {\em Nuclear Fusion}, 38(12):1881--1891, dec 1998.

\bibitem{de_baar_99}
M.~R. de~Baar, M.~N.~A. Beurskens, G.~M.~D. Hogeweij, and N.~J. Lopes~Cardozo.
\newblock Tokamak plasmas with dominant electron cyclotron heating; evidence
  for electron thermal transport barriers.
\newblock {\em Physics of Plasmas}, 6(12):4645--4657, 1999.

\bibitem{Schilham_2001}
A~M~R Schilham, G~M~D Hogeweij, and N~J~Lopes Cardozo.
\newblock Electron thermal transport barriers in {RTP}: experiment and
  modelling.
\newblock {\em Plasma Physics and Controlled Fusion}, 43(12):1699--1721, nov
  2001.

\bibitem{DeBaar2005}
M.~R. {De Baar}, A.~Thyagaraja, G.~M.D. Hogeweij, P.~J. Knight, and E.~Min.
\newblock {Global plasma turbulence simulations of q = 3 sawtoothlike events in
  the RTP tokamak}.
\newblock {\em Physical Review Letters}, 94(3):6--9, 2005.

\bibitem{Mantica2005}
P.~Mantica, A.~Thyagaraja, J.~Weiland, G.~M.D. Hogeweij, and P.~J. Knight.
\newblock {Heat pinches in electron-heated Tokamak plasmas: Theoretical
  turbulence models versus experiments}.
\newblock {\em Physical Review Letters}, 95(18):1--4, 2005.

\bibitem{Petty_1994}
C.C. Petty and T.C. Luce.
\newblock Inward transport of energy during off-axis heating on the diii-d
  tokamak.
\newblock {\em Nuclear Fusion}, 34(1):121, jan 1994.

\bibitem{Smith_2015}
S.P. Smith, C.C. Petty, A.E. White, C.~Holland, R.~Bravenec, M.E. Austin,
  L.~Zeng, and O.~Meneghini.
\newblock Electron temperature critical gradient and transport stiffness in
  diii-d.
\newblock {\em Nuclear Fusion}, 55(8):083011, jul 2015.

\bibitem{TS_ref}
T.~N. Carlstrom, G.~L. Campbell, J.~C. DeBoo, R.~Evanko, J.~Evans, C.~M.
  Greenfield, J.~Haskovec, C.~L. Hsieh, E.~McKee, R.~T. Snider, R.~Stockdale,
  P.~K. Trost, and M.~P. Thomas.
\newblock Design and operation of the multipulse thomson scattering diagnostic
  on diii‐d (invited).
\newblock {\em Review of Scientific Instruments}, 63(10):4901--4906, 1992.

\bibitem{ECE_ref}
M~E Austin and J~Lohr.
\newblock {Electron cyclotron emission radiometer upgrade on the DIII-D
  tokamak}.
\newblock {\em Review of Scientific Instruments}, 74(3):1457--1459, 2003.

\bibitem{CER_ref}
P~Gohil, K~H Burrell, R~J Groebner, and R~P Seraydarian.
\newblock {High spatial and temporal resolution visible spectroscopy of the
  plasma edge in DIII‐D}.
\newblock {\em Review of Scientific Instruments}, 61(10):2949--2951, 1990.

\bibitem{TORAY_ref}
K~Matsuda.
\newblock {Ray tracing study of the electron cyclotron current drive in DIII-D
  using 60 GHz}.
\newblock {\em IEEE Transactions on Plasma Science}, 17(1):6--11, 1989.

\bibitem{Lao_1985_EFIT}
L~L Lao, H~St. John, R~D Stambaugh, A~G Kellman, and W~Pfeiffer.
\newblock {Reconstruction of current profile parameters and plasma shapes in
  tokamaks}.
\newblock {\em Nuclear Fusion}, 25(11):1611--1622, nov 1985.

\bibitem{TRANSP_ref}
Joshua Breslau, Marina Gorelenkova, Francesca Poli, Jai Sachdev, Alexei Pankin,
  Gopan Perumpilly, Xingqiu Yuan, and Laszlo Glant.
\newblock Transp.
\newblock Computer Software, jun 2018.

\bibitem{TRANSP_OMFIT_ref}
B~A Grierson, X~Yuan, M~Gorelenkova, S~Kaye, N~C Logan, O~Meneghini, S~R
  Haskey, J~Buchanan, M~Fitzgerald, S~P Smith, L~Cui, R~V Budny, and F~M Poli.
\newblock {Orchestrating TRANSP Simulations for Interpretative and Predictive
  Tokamak Modeling with OMFIT}.
\newblock {\em Fusion Science and Technology}, 74(1-2):101--115, 2018.

\bibitem{MSE_ref}
B~W Rice, D~G Nilson, and D~Wr{\'{o}}blewski.
\newblock {Motional Stark effect upgrades on DIII‐D}.
\newblock {\em Review of Scientific Instruments}, 66(1):373--375, 1995.

\bibitem{Harvey_non_thermal}
R~W Harvey, M~R O'Brien, V~V Rozhdestvensky, T~C Luce, M~G McCoy, and G~D
  Kerbel.
\newblock {Electron cyclotron emission from nonthermal tokamak plasmas}.
\newblock {\em Physics of Fluids B: Plasma Physics}, 5(2):446--456, 1993.

\bibitem{Subhas_non_thermal}
P~V Subhash, Amit~Kumar Singh, Hitesh Pandya, V~S Divya, M~P Aparna, and
  T~K~Basitha Thanseem.
\newblock {A Parametric Model for Contribution of Superthermal Electrons to
  Oblique Measurement Electron Cyclotron Spectra Under ITER-Like Conditions}.
\newblock {\em Fusion Science and Technology}, 72(1):49--59, 2017.

\bibitem{ONETWO_ref}
H.~St.~John, T.~Taylor, Y.~Lin-Liu, and A.~Turnbull.
\newblock Transport simulation of negative magnetic shear discharges.
\newblock In {\em Plasma Physics and Controlled Nuclear Fusion Research},
  volume~3, page 603, 1994.

\bibitem{Mantica2000}
P.~Mantica, G.~Gorini, G.~M.D. Hogeweij, N.~J.~Lopes Cardozo, and A.~M.R.
  Schilham.
\newblock Heat convection and transport barriers in low-magnetic-shear
  rijnhuizen tokamak project plasmas.
\newblock {\em Physical Review Letters}, 85:4534--4537, 2000.

\bibitem{jacchia_cylinder_chi}
A.~Jacchia, P.~Mantica, F.~De~Luca, and G.~Gorini.
\newblock Determination of diffusive and nondiffusive transport in modulation
  experiments in plasmas.
\newblock {\em Physics of Fluids B: Plasma Physics}, 3(11):3033--3040, 1991.

\bibitem{Schmitz_2012}
L.~Schmitz, C.~Holland, T.L. Rhodes, G.~Wang, L.~Zeng, A.E. White, J.C.
  Hillesheim, W.A. Peebles, S.P. Smith, R.~Prater, G.R. McKee, Z.~Yan, W.M.
  Solomon, K.H. Burrell, C.T. Holcomb, E.J. Doyle, J.C. DeBoo, M.E. Austin,
  J.S. deGrassie, and C.C. Petty.
\newblock Reduced electron thermal transport in low collisionality h-mode
  plasmas in diii-d and the importance of tem/etg-scale turbulence.
\newblock {\em Nuclear Fusion}, 52(2):023003, jan 2012.

\bibitem{kobayashi_2023}
T.~Kobayashi, Z.~Yan, G.~R. McKee, M.~E. Austin, B.~A. Grierson, and P.~Gohil.
\newblock {Prompt core confinement improvement across the L–H transition in
  DIII-D: Profile stiffness, turbulence dynamics, and isotope effect}.
\newblock {\em Physics of Plasmas}, 30(3):032301, 03 2023.

\bibitem{Austin2006}
M.~E. Austin, K.~H. Burrell, R.~E. Waltz, K.~W. Gentle, P.~Gohil, C.~M.
  Greenfield, R.~J. Groebner, W.~W. Heidbrink, Y.~Luo, J.~E. Kinsey, M.~A.
  Makowski, G.~R. McKee, R.~Nazikian, C.~C. Petty, R.~Prater, T.~L. Rhodes,
  M.~W. Shafer, and M.~A.~Van Zeeland.
\newblock Core barrier formation near integer q surfaces in diii-d.
\newblock {\em Physics of Plasmas}, 13:1--8, 2006.

\bibitem{Osorio2021_ITB_q}
L.~A. Osorio, M.~Roberto, I.~L. Caldas, R.~L. Viana, and Y.~Elskens.
\newblock Onset of internal transport barriers in tokamaks.
\newblock {\em Physics of Plasmas}, 28, 2021.

\bibitem{cold_pulse}
P.~Rodriguez-Fernandez, A.~E. White, N.~T. Howard, B.~A. Grierson, G.~M.
  Staebler, J.~E. Rice, X.~Yuan, N.~M. Cao, A.~J. Creely, M.~J. Greenwald,
  A.~E. Hubbard, J.~W. Hughes, J.~H. Irby, and F.~Sciortino.
\newblock Explaining cold-pulse dynamics in tokamak plasmas using local
  turbulent transport models.
\newblock {\em Physical Review Letters}, 120:75001, 2018.

\end{thebibliography}

\end{document}